\documentclass[11pt]{article}
\usepackage{amsmath,amsfonts, amssymb, braket, bbold}
\usepackage{tikz}
\usetikzlibrary{trees,er,snakes,shapes,mindmap}
\usepackage{titlesec}
\titleformat{\section}
 {\normalfont\fontfamily{put}\fontsize{12pt}{16pt}\bfseries\color{black}}
{\thesection}{1em}{}
\def \beq  {\begin{equation}}
\def \eeq  {\end{equation}}
\def \beqar {\begin{eqnarray}}
\def \eeqar {\end{eqnarray}}
\allowdisplaybreaks
\def\sqr#1#2{{\vcenter{\vbox{\hrule height.#2pt
\hbox{\vrule width.#2pt height#1pt \kern#1pt
\vrule width.#2pt}\hrule height.#2pt}}}}

\def\L {{\cal L}}

\def\vf {{\varphi}}

\def\Tr {{\rm Tr}}

\def\del {\partial}

\def\a {\alpha}

\def\l {\lambda}

\def\bz {{\bar{z}}}

\def\A {{\cal A}}

\def\D {{\cal D}}

\def\H {{\cal H}}
\def\I{{\cal I}}
\def \L {{\cal L}}
\def\M{{\cal M}}

\def\vf {{\varphi}}

\def\half{\textstyle{1\over 2}}

\mathchardef\mhyphen="2D
\begin{document}
%%%%%%%%%%%%%%%%%%%%%%%%%%%%%%%%%%%%%%%%%%%%%%%
%%%%%%%%%%%%%%%%%%%%%%%%%%%%%%%%%%%%%%%%%%%%%%%
%\fontfamily{pnb}\fontsize{12pt}{16pt}\selectfont
%\fontfamily{pzc}\fontsize{14pt}{16pt}\selectfont
%\fontfamily{pbk}\fontsize{12pt}{16pt}\selectfont
%\fontfamily{cmr}\fontsize{11pt}{15pt}\selectfont
\fontfamily{put}\fontsize{12pt}{17pt}\selectfont
%\fontfamily{lmss}\fontsize{11pt}{16pt}\selectfont
%\fontfamily{phv}\fontshape{ro}\fontsize{11pt}{14pt}\selectfont
%\fontfamily{ptm}\fontseries{m}\fontshape{r}\fontsize{12pt}{16pt}\selectfont
%\fontfamily{pnc}\fontseries{m}\fontshape{r}\fontsize{11pt}{15pt}\selectfont
%\fontfamily{ppl}\fontseries{m}\fontshape{r}\fontsize{11pt}{15pt}\selectfont
%\usefont{T1}{phv}{m}{it}
%%%%%%%%%%%%%%%%%%%%%%%%%%%%%%%%%%%%%%%%%%%%%%%
%%%%%%%%%%%%%%%%%%%%%%%%%%%%%%%%%%%%%%%%%%%%%%%
\def \CMP {{Commun. Math. Phys.}}
\def \PRL {{Phys. Rev. Lett.}}
\def \PL {{Phys. Lett.}}
\def \NPBProc {{Nucl. Phys. B (Proc. Suppl.)}}
\def \NP {{Nucl. Phys.}}
\def \RMP {{Rev. Mod. Phys.}}
\def \JGP {{J. Geom. Phys.}}
\def \CQG {{Class. Quant. Grav.}}
\def \MPL {{Mod. Phys. Lett.}}
\def \IJMP {{ Int. J. Mod. Phys.}}
\def \JHEP {{JHEP}}
\def \PR {{Phys. Rev.}}
\def \JMP {{J. Math. Phys.}}
\def \GRG{{Gen. Rel. Grav.}}
%%%%%%%%%%%%%%%%%%%%%%%%%%%%%%%%%%%%%%%%%%%%%%%
%%%%%%%%%%%%%%%%%%%%%%%%%%%%%%%%%%%%%%%%%%%%%%%
\begin{titlepage}
\null\vspace{-62pt} \pagestyle{empty}
\begin{center}
%\rightline{CCNY-HEP-18/4}
%\rightline{August 2018}
\vspace{1.3truein} {\Large\bfseries
Matter-gravity coupling for fuzzy geometry and}\\
\vskip .15in
{\Large\bfseries the Landau-Hall problem }\\
\vskip .5in
{\Large\bfseries ~}\\
%%%%%%%%%%%%%%%%%%%%%%%%%%%%%%%%%%%%%%%%%%%%%%%
%%%%%%%%%%%%%%%%%%%%%%%%%%%%%%%%%%%%%%%%%%%%%%%
{\large\sc V.P. Nair}\\
\vskip .2in
{\sl Physics Department,
City College of the CUNY\\
New York, NY 10031}\\
 \vskip .1in
\begin{tabular}{r l}
{\sl E-mail}:&\!\!\!{\fontfamily{cmtt}\fontsize{11pt}{15pt}\selectfont vpnair@ccny.cuny.edu}\\
\end{tabular}
\vskip 1.5in
%%%%%%%%%%%%%%%%%%%%%%%%%%%%%%%%%%%%%%%%%%%%%%%
%%%%%%%%%%%%%%%%%%%%%%%%%%%%%%%%%%%%%%%%%%%%%%%
\centerline{\large\bf Abstract}
\end{center}
We consider a set of physical degrees of freedom coupled to a finite-dimensional Hilbert space,
which may be taken as modeling a fuzzy space or as the lowest Landau level of
a Landau-Hall problem. These may be viewed as matter fields on a fuzzy space.
Sequentially generalizing to arbitrary backgrounds, we argue that
the effective action is given by the Chern-Simons form associated with the Dirac index
density (with gauge and gravitational fields), with an abelian gauge field 
shifted by the Poincar\'e-Cartan form for matter dynamics.
The result is an action for matter fields where
the Lagrangian is integrated with a density which is a specific polynomial
in the curvatures.
\end{titlepage}
%%%%%%%%%%%%%%%%%%%%%%%%%%%%%%%%%%%%%%%%%%%%%%%
%%%%%%%%%%%%%%%%%%%%%%%%%%%%%%%%%%%%%%%%%%%%%%%
\fontfamily{put}\fontsize{12pt}{17pt}\selectfont
\pagestyle{plain} \setcounter{page}{2}
\section{Introduction}

The energy levels of a charged particle in a magnetic field, the so-called Landau levels, have long been a useful structure to analyze many questions of physical interest. The quantum Hall effect is perhaps the most direct example of the use of these Landau levels \cite{generalQHE}.
In this context, a number of variants, including different topologies and different geometries
(as characterized by metrics and spin connection) \cite{{QHE1},{QHE1a}}, as well as higher dimensions \cite{{QHE2}, {QHE3}, {KN-1}, {KN-rev}}
have also been explored.
The Landau levels have also provided a useful analytical tool for discussing effective actions,
pair production by both Abelian and nonabelian gauge fields, etc. \cite{QHEother}.
Another important 
reason for research interest in this area has to do with noncommutative geometry \cite{NCgeneral}.
The set of degenerate states of a fixed Landau level can be used as a model for a noncommutative
manifold, with operators on these degenerate states providing observables for the
noncommutative space.
The existence of symbols and star-products corresponding to such operators
render the continuum or commutative space approximation to such noncommutative spaces
easily tractable. 
It is worth emphasizing  that
noncommutative geometry has been a recurrent paradigm for many approaches to 
quantum gravity, both intrinsically as an idea in its own right \cite{NCgeneral}
and as special cases in string theory \cite{NCstring}.
Needless to say, there has been an enormous amount of recent research along these lines.

Offset against this large body of literature, it is interesting that there are still many
open questions of physical relevance.
If we consider the LLL as a model for a noncommutative space, we can construct fields living
on such spaces. What are the characteristics of such a field theory? 
This 
is the key question we analyze here.
The construction of noncommutative field theories has a long history in its own right. 
Most of this work has been
at the level of promoting products of fields and their spatial (or spacetime) derivatives to star
products, but using standard Lagrangians \cite{NCfield}. We are considering the construction 
of the action starting from operators on the Hilbert space (i.e., LLL) modeling the
noncommutative space. The resulting action will be different in many features, especially 
in its relation to the background geometry. We have argued elsewhere
that the LLL analysis can be used for understanding gravity on noncommutative spaces
\cite{VPN1}.
The present work may be viewed as extending such ideas to include matter 
couplings to gravity. Phrased another way, we ask the question whether there are
particular peculiarities for matter-gravity coupling which we can extract from
analyzing fields acting on the LLL.

Since we are modeling the noncommutative space by the LLL,
there is a possibility of some confusion about the roles of the fields we are discussing.
It is useful to clarify this at the outset. We will consider degrees of freedom which eventually
lead to a set of fields we  shall refer to as ``matter fields", designated by
$\phi$. But there will be a set of fermion fields defining the LLL itself , i.e., the
noncommutative geometry. These latter ones, which we designate by 
$\psi$, $\psi^\dagger$, are what we shall refer to as the ``spatial fields".
The question of interest for us is how the dynamics for the $\phi$-fields is affected
by the background geometry for $\psi$, $\psi^\dagger$.
This is not simply a matter of writing an coupled action for both sets of fields and 
analyzing it, as we would normally do for interacting field theories, because, for us, the $\phi$-fields do not exist outside of the LLL. This is the distinctive feature of our analysis.

While the noncommutative geomtery-gravity angle is the
natural setting of the problem, it
may also be viewed as a much more standard physical 
problem, of interest
within the quantum Hall setting.
 If a set of fields $\phi$
are coupled to 
charged fermions (described by $\psi$, $\psi^\dagger$)
and if these fermions are confined to the LLL, what is the theory of the fields $\phi$
within the LLL? How does this field theory respond to changes in the background fields, metric and spin connection? Clearly this is a natural next step to the many analyses which have been done for the pure electron system with arbitrary background gauge fields and metrics \cite{{QHE1},{QHE1a}}.

The organization of the paper and overall flow of logic 
may be summarized as follows.
 We start with the dynamics of a physical system whose
observables are matrices acting on the states of the LLL (or the Hilbert space modeling the
spatial geometry). Not surprisingly, this leads to a Hamiltonian or Lagrangian with star
products for the fields and their derivatives. 
We will consider the required mathematical framework for the two-dimensional case in section 2,
 the more general higher dimensional cases in section 3.
Complex K\"ahler  geometry will play a crucial role in defining the star products.
The path-integral for the dynamics of the physical system under consideration,
which we designate the matter fields, we will argue,
is defined by a Chern-Simons action (related to the Dolbeault index density)
with
a shift of the (abelian) gauge potential by the Poincar\'e-Cartan form.
So far, the results will still be tied to the complex geometry
of the background.
Next, in section 4,
we want to generalize this to more general gauge and gravitational backgrounds.
Towards this, we argue that there is a scaling of coordinates under which, if we restrict to
low energy physics, one can ignore higher terms in the  star products,
 thus giving an approximation not tied to the complex geometry.
The resulting version can then be embedded in a more general geometry
and the effective action constructed in terms of the Chern-Simons form associated with
the Dirac index density. Again, the prescription for matter fields is to shift the abelian gauge field 
in the Chern-Simons form for the Dirac index by
the Poincar\'e-Cartan form for matter fields.
Explicit formulae for the effective action in 2+1 and 4+1 dimensions are given.
Finally, we give an action for a set of fermion fields, to be viewed as the fields which
eventually define
the spatial manifold, which leads to the prescription for the matter couplings
we have obtained.
The paper concludes with a short summary/discussion.

Explicit derivations of the effective action with
perturbations to the background geometry and gauge fields for the LLL, in the absence of what
we have referred to as matter fields,
were given in \cite{{QHE1}, {QHE1a}, {KN-1}} in 2+1 dimensions and
in \cite{{KN-1}, {KN-rev}} for higher dimensions.
Also, a different way of constructing an effective action for the Landau-Hall problem for all odd spacetime dimensions, using the Dolbeault index theorem,
was given  in \cite{KN-Dolb}.
The present work may be considered as an extension of these works
 to include matter couplings,
 and also to accommodate more general, not  necessarily complex, geometries.
An interesting feature of the emergent matter-gravity coupling is that the action is given by
integrating the matter Lagrangian with
a density which is a specific polynomial involving
powers of the curvature. It is interesting to note that such
couplings for matter and gravity have been the subject of recent research motivated by
issues with dark matter \cite{dark}.

%%%%%%%%%%%%%%%%%%%%%%%%%%%%%%%%%%%%%%%%%%%%%%%
%%%%%%%%%%%%%%%%%%%%%%%%%%%%%%%%%%%%%%%%%%%%%%%
\section{Matter fields and gravity and the LLL in two spatial dimensions}
%%%%%%%%%%%%%%%%%%%%%%%%%%%%%%%%%%%%%%%%%%%%%%%
%%%%%%%%%%%%%%%%%%%%%%%%%%%%%%%%%%%%%%%%%%%%%%%
We start by considering a physical system characterized by a set of operators
which are the relevant dynamical
variables. Among the operators, we assume there is a mutually commuting set
which we denote by $\{ q_\l \} $, where $\l$ is an index labeling the distinct operators.
Since we are aiming for a field theory eventually, we take the
eigenvalues of the $q$'s to form a continuous set.
The states of the physical system can be taken as the vectors
$\ket{q}$ in a Hilbert space $\H$.
Any nontrivial dynamics should allow for altering the state of the system, so there
must be operators which do not commute with
the $q$'s. We  can take them to be a set of conjugate variables
$\{ p_\l \}$. Taking the Hamiltonian to be a function of these variables
$\{ q_\l, p_\l \}$, time-evolution of the system by an infinitesimal amount $\epsilon$
is described by the transformation kernel
\beq
\bra{q'} e^{- i H \epsilon} \ket{q} = \int [dp] \, \exp\left[ i p_\l (q'_\l - q_\l ) - H(p,q) \, \epsilon \right]
\label{pint1}
\eeq
It is also
possible to carry  out the integration over the $p$'s and write this in terms of the action.

The key point for us is that we want to interpret this as a field theory
in the  language of noncommutative geometry.
The variables $\{ q_\l \}$ should describe a field operator on some manifold $\M$
in a suitable large $N$ limit. For this consider an $N$-dimensional Hilbert space
$\H_N$. This is not the Hilbert space $\H$ of the physical system we are considering,
but  the sequence of $\H_N$'s will model the
noncommutative version of $\M$.
Observables on $\H_N$ correspond to $N \times N$ matrices. Thus we want to
identify $q_\l$ as the mode amplitudes for a matrix ${\hat q}$, with matrix elements
${\hat q} _{ij}$, expanded as
\beq
{\hat q}_{ij} = \sum_\l q_\l \, (T_\l )_{ij}, \hskip .3in
i, j = 1, 2, \cdots, N,
\label{pint2}
\eeq
where $\{ T_\l \}$ form a basis for $N \times N$ matrices.
We can take this to be an orthonormal basis obeying
$\Tr (T_\l T_{\l'}) = \delta_{\l \l'}$.
In the large $N$ limit, the algebra of the $N\times N$ matrices should become the algebra of functions
on $\M$, with
$T_\l$ corresponding to a complete set of mode functions.
There are two ways to pass from $\{q_\l\}, \{p_\l\}$ to functions on $\M$.
If $\M$ is a compact K\"ahler manifold, which is mostly the case we will be considering,
we can
take a suitable multiple of the K\"ahler form as a symplectic structure and carry out quantization
to construct $\H_N$. In this case, there will be a set of orthonormal ``wave functions"
$u_i$ which are holomorphic. Strictly speaking, these are sections of a suitable power of the canonical line bundle on $\M$. The set $\{ u_i \}$ can also be viewed as coherent states on $\M$ obtained via
standard coherent state quantization. Given this structure, there is a function $\phi$
on $\M$ such that the matrix elements ${\hat q}_{ij}$ in (\ref{pint2}) can be obtained as
\beq
{\hat q}_{ij} = \int d\mu \, u_i^* \, \phi \, u_j
\label{pint3}
\eeq
The function $\phi$ is the contravariant symbol for
${\hat q}_{ij}$ and the prescription (\ref{pint3}) is the Berezin-Toeplitz (BT) quantization
of $\phi$.

If ${\hat A}$, ${\hat B}$ are $N \times N$ matrices, then the function which gives
the operator or matrix product $({\hat A} {\hat B})_{ij}$
via (\ref{pint3}) is the star-product of the functions $A$ and $B$
corresponding to the individual matrices; i.e,
\beq
({\hat A} {\hat B})_{ij} = \int d\mu~ u_i^* ( A * B ) \, u_j
\label{pint4}
\eeq
The trace of a matrix ${\hat A}$ can be written as
\beq
\Tr ({\hat A} ) = \sum_i {\hat A}_{ii} = \int d\mu~\Bigl[  \sum_i u^*_i u_i \Bigr] \,
A = \int d\mu~ \rho~ A
\label{pint5}
\eeq
We see that $\rho = \sum_i u^*_i u_i$ defines a density to be used in the integration.

Using these formulae, we can convert terms in $H$ (and the action) into integrals over
the star-products of various contravariant symbols. Thus
\beq
\sum_\lambda p_\l p_\l = \sum_{\l, \l'} \Tr ( p_\l T_\l ) (p_{\l'} T_{\l'} ) = \Tr ( {\hat \Pi}\, {\hat \Pi})
= \int d\mu\, \rho~ \Pi* \Pi
\label{pint6}
\eeq
where ${\hat \Pi}_{ij}  = \sum_\l p_\l (T_\l)_{ij}$. As an example, consider choosing a Hamiltonian
of the form
\beq
H = {1\over 2} \Tr \left[ {\hat \Pi} \,{\hat \Pi}  + \beta_1\, [T_\a, {\hat q}]\, [ T_\a, {\hat q}] + m_0^2 {\hat q} {\hat q}
\right] + g_0\, \Tr ( {\hat q}^4)
\label{pint7}
\eeq
where $\beta_1$, $m_0$ and $g_0$ are arbitrary constants.
The last two may be identified as the bare mass and bare coupling constant.
The commutator $[T_\a , {\hat q}]$ stands for the matrix version of the derivative,
$T_\a$ being a specific set of matrices. 
Since we have not specified exactly how the commutators translate to derivatives and since we may
have to do some scaling of spatial coordinates, we must allow for an arbitrary
coefficient $\beta_1$ for the $[T_\alpha , {\hat q}]^2$-term. 
Expression (\ref{pint7}) leads to the
field theory Hamiltonian
\beq
H (\Pi, \phi)  = \int d\mu\, \rho~ \left[ {1\over 2} \left( \Pi * \Pi + \alpha_1 \,(\nabla_\a\phi)* (\nabla_\a\phi) 
+ m_0^2 \phi *\phi \right) + g_0\, \phi* \phi* \phi* \phi \right]
\label{pint8}
\eeq
Here $\alpha_1$ is the version of $\beta_1$ once we make the translation of the commutator
$[T_\a, {\hat q}]$ to a derivative of the field.
If star products are approximated by ordinary products, which may be reasonable
 as
$N \rightarrow \infty$, then we get a familiar form of the
Hamiltonian density integrated with $d\mu\, \rho$ as the
volume element.

Returning to the
transformation kernel in  (\ref{pint1}), we first use the product of a sequence of such kernels and integrate over the $q$'s to obtain the
Hamiltonian path integral in the usual way,
\beqar
Z &=&{\cal N} \int [Dp \, Dq] \, \exp \left( i \int dt \, [p_\l {\dot q}_\l  - H (p, q)  ] \right)\nonumber\\
&=& \int [Dp \, Dq] \, \exp \left( - \int \A (p, q)   \right)\nonumber\\
\A&=& -i \left[ p_\l {\dot q}_\l  - H (p, q) \right]\, dt
\label{pint8a}
\eeqar
Here $\A$ is the Poincar\'e-Cartan form for the system under consideration
and ${\cal N}$ is a normalization factor.
(We define  $\A$ to be antihermitian to agree with the convention used later for the
gauge fields.)
Written in terms of symbols, this expression for the path integral reads
\beq
Z = {\cal N} \int [D\Pi\, D\phi]\, \exp\left( - \int d\mu \, \rho~ \A \right)
\label{pint8b}
\eeq
where we now have the symbol for $\A$, also written as $\A$,
 in the exponent. Rewriting this in terms of the individual symbols for
 $p$ and $q$ would require the star products. Thus we can also write
\beq
Z = {\cal N} \int [D \Pi\, D\phi]\, \exp\left( i \int dt\,d\mu\, \rho~ \left[\Pi* {\dot \phi}  - H (\Pi, \phi )\right]
\right)
\label{pint8c}
\eeq 

The second way of passing from matrices to functions is via the covariant symbol. Here we start from the matrix elements of an operator ${\hat A}_{ij}$ and form a function  $(A)$ defined by
\beq
(A) = \sum_{i j} \D_i \, {\hat A}_{ij} \D^*_j , \hskip .3in
\D_i = {u_i \over \sqrt{N}}
\label{pint9}
\eeq
Notice that the covariant symbol in the above equation
defines a function on $\M$ given the
matrix elements ${\hat A}_{ij}$, while the contravariant symbol is a function on
$\M$ which leads to the matrix elements via (\ref{pint3}).
In this sense, they are converses of each other, but the symbols are not identical in general.
By appropriately using the completeness properties of the $\D$'s, one can again pass from
a Hamiltonian as in (\ref{pint7}) to the form $(\ref{pint8})$, with $\Pi, \, \phi$
replaced by the covariant symbols $(\Pi)$, $(\phi)$ and the star product should also
be the one pertaining to the covariant symbols. While this method has been used in a number of applications (for example, see \cite{ {KN-1}, {KN-rev}}), for what follows, we shall mostly use the contravariant symbols, although we will give a more explicit formula
for the covariant symbol for $\M = S^2$ later.

The passage from a matrix expression to functions (with star products) as in (\ref{pint8}) has been
well known for many years. But our aim here is to go beyond that and consider the situation where there are perturbations to the background gauge fields and the spin connection
on $\M$, these being the data needed for constructing $\H_N$ and $u_i$.

Secondly, $\H_N$ and $u_i$ are obtained as the lowest Landau level (LLL) and the corresponding set of wave functions for a Landau-Hall problem on $\M$. So we can apply the analysis to various fields
coupled to the electrons which fill the LLL.
(We are taking the fields to refer to observables restricted to the
LLL only. Fields which have an existence outside of the Hall system will have additional terms
in the Hamiltonian and the action.)

There is another reason why the  LLL setting is useful.
Given the $N$-dimensional vector space $\H_N$ and matrices as linear transformations of 
$\H_N$, we need the $u_i$ to define symbols and star products.
This choice is not unique. Hence the large $N$ limit we obtain can be different
for different choices. 
For example, the continuum limit may correspond to the symplectic structure
$n\, \Omega_K$ (where $\Omega_K$ is the K\"ahler form) or a perturbation of it in the form
$n\, \Omega_K + d (\delta \a )$ since both will lead to the same number of states, at least for large
$n$. This is equivalent to different choices of the background gauge fields.
Is there an optimal choice? This would require a criterion selecting a particular 
background (of  gauge fields and geometry) and so it would be the key principle for
gravity on noncommutative spaces \cite{VPN1}. The placement of the problem in the LLL context 
gives a simple calculational scheme to analyze such questions.

%%%%%%%%%%%%%%%%%%%%%%%%%%%%%%%%%%%%%%%%%%%%%%%
%%%%%%%%%%%%%%%%%%%%%%%%%%%%%%%%%%%%%%%%%%%%%%%
\section{Matter fields and gravity and the LLL in higher dimensions}
%%%%%%%%%%%%%%%%%%%%%%%%%%%%%%%%%%%%%%%%%%%%%%%
%%%%%%%%%%%%%%%%%%%%%%%%%%%%%%%%%%%%%%%%%%%%%%%

We start with the framework for the
Landau levels and the set-up of  the $\nu =1$ state.
For this we consider fields
$\psi$, $\psi^\dagger$ which represent the electron or the charged fermions.
They are subject to a $U(1)$ background gauge field, i.e., the magnetic field,
and we will consider the fully filled lowest Landau level (LLL) for these,
i.e., the $\nu = 1$ Hall state.
From the point of view of noncommutative geometry, the
LLL will define the Hilbert space $\H_N$ which serves as
the model for the noncommutative version of $\M$.
Thus the fields $\psi$, $\psi^\dagger$ define the noncommutative
spatial geometry.
For this reason, and to avoid confusion with the fields $\phi$ introduced previously,
we will refer to $\psi$, $\psi^\dagger$ as the spatial fields.
The set of fields $\phi$ will be referred to as matter fields.

Towards setting up the Landau levels and the $\nu =1$ state,
initially we will consider the spatial manifold to be
$S^2$, so that spacetime is $S^2 \times {\mathbb R}$ \cite{{haldane},{QHE2}}.
The background magnetic field which leads to the Landau levels
is thus a uniform magnetic field on $S^2$, corresponding to a magnetic
monopole at the center if we consider the $S^2$ as embedded in three
dimensions.
The Hamiltonian for the $\psi, \psi^\dagger$ fields has the form
\beq
H = \int d\mu(g)\, \psi^\dagger \left[ {R_+ R_- \over 2 m r^2}\right] \psi
\label{matt1}
\eeq
We will view $S^2$ as $SU(2)/U(1)$, so that we can use an
$SU(2)$ group element $g$ to coordinatize
the spatial manifold, modulo the $U(1)$ identification.
 On this group element, viewed as a $2\times 2$ matrix,
one can define left and right translation operators via
\beq
L_a \, g = t_a \, g, \hskip .3in
R_a \, g = g \, t_a
\label{matt2}
\eeq
where $t_a$ are a basis for the generators of $SU(2)$ in the
$2\times 2$ matrix representation. They may be taken as
$t_a= {\half} \sigma_a$, where $\sigma_a$ are the Pauli matrices.
The operators $R_\pm = R_1 \pm i R_2$  appearing in the
Hamiltonian (\ref{matt1}), are thus translation operators
on $S^2$. Also $d\mu (g)$ denotes the Haar measure on
$SU(2)$ with the normalization
$\int d\mu = 1$. The volume on $S^2$ differs from the $SU(2)$ volume
by the $U(1)$ factor. Since we will be considering integrands which are invariant
under the $U(1)$ action, integration over  this  extra $U(1)$ is immaterial and
we will use the full $SU(2)$ volume. 
$r$ is a scale parameter, which may be viewed as the radius of $S^2$.

The translation operators $R_\pm$ can be identified as
covariant derivatives, so that
having a nonzero background magnetic field $B$ is equivalent to
the requirement that the fermion fields obey
$[R_+, R_-] \psi = 2 R_3\psi  = - n\psi  = - 2 B r^2\psi $. Here $n$ is an integer
in accordance with the Dirac quantization condition.
The eigenmodes of $R_+ R_-$ take the form
\beq
U^{(q)}_A = \sqrt{2 q + n +1}~ \D^{({n\over 2}+q)}_{A, - {n\over 2}}(g)
\label{matt3}
\eeq
where $\D^{(j)}_{A, B} (g)$ are the representatives of 
$g$ in the spin-$j$ representation with $A, B$ labeling states within the representation.
They take values
$1, 2, \cdots, (2 j+1)$. Further, $q$ is a semi-positive integer, with
$q = 0$ corresponding to the lowest Landau level.
The fermion field $\psi$ has the mode expansion
\beq
\psi = \sum_i a_i \, u_i + \sum_{q\neq 0} a^{(q)}_A \, U^{(q)}_A
\label{matt4}
\eeq
where we have separated out the LLL, with $u_i = \sqrt{n+1}\, \D^{({n\over 2})}_{i, -{n \over 2}}$.
In terms of the
creation and annihilation operators $a_i$, $a^\dagger_i$,
which obey the standard fermion algebra,
the completely filled LLL state is given by
\beq
\ket{w} = a_1^\dagger\,a_2^\dagger\,\cdots a_N^\dagger \ket{0}
\label{matt5}
\eeq
with $N = n+1$.

There are $N$ single particle states corresponding to the LLL.
These are characterized by the wave functions $u_i$ in  (\ref{matt4}).
They form the basis for a one-particle Hilbert space
$\H_N$, which models fuzzy $S^2$.
They can also be constructed directly without embedding them in the larger framework
of Landau levels.
In terms of the group element $g$,
the K\"ahler forms are given by
\beq
\a_K= i \Tr (t_3 g^{-1} dg ) , \hskip .3in
\Omega_K = d \a_K = -i \Tr ( t_3 g^{-1} dg \, g^{-1} dg )
\label{matt5a}
\eeq
These define the canonical line bundle for $S^2 \sim {\mathbb{CP}^1}$.
The $n$-th power of the canonical line bundle has
the curvature $ \Omega = n\, \Omega_K$ and $u_i$ are sections of this line bundle.
They are holomorphic since they obey
\beq
R_- u_i = \sqrt{N} \, R_- \D^{({n\over 2})}_{i, - {n \over 2}} = 0
\label{matt5b}
\eeq
These are the coherent states obtained by straightforward quantization 
of $(S^2, n \Omega_K )$ with
the holomorphic polarization.

Observables restricted to the LLL are $N \times  N$ matrices, which, as mentioned
in section 2,
can be expanded in terms of the basis
$\{ T_\l \}$.
Since we are considering
$S^2$, such a basis is provided by
the matrix analogs of the spherical harmonics.
These are given by
\beq
\{ T_\lambda \} = \Big\{ {1\over \sqrt{N}}, {T_a\over \sqrt{3 j (j+1) (2 j +1)}}, \cdots
\Big\}
\label{matt10}
\eeq
with $j = {n \over 2}$.
Thus we have a series of tensor operators $T_\lambda$
with $SU(2)$ angular momentum $l =
0, 1, \cdots n$. The series naturally terminates at $l =n$ for 
$N \times N$ matrices. We have chosen the normalization
condition $\Tr (T_\lambda T_{\lambda'}) = \delta_{\lambda \lambda'}$.
The symbols corresponding to these matrices
become the usual spherical harmonics as $n \rightarrow \infty$.
In this way, the space of functions on the LLL lead to the commutative algebra
of functions on $S^2$ as $n \rightarrow \infty$, in accordance with 
the expected structure for fuzzy $S^2$.

For this example of the fuzzy version of $S^2$, we can specify the
covariant symbol for ${\hat A}$ more explicitly as
\beq
({\hat A}) = \sum_{i k} \D^{({n\over 2})}_{i, - {n \over 2}}(g) \, 
{A}_{ik} \, \D^{({n\over 2})*}_{k, - {n \over 2}}(g) 
\label{matt13}
\eeq
This is clearly a function on $S^2$. Notice that the normalized wave functions are
$u_i = \sqrt{N}\, \D^{({n\over 2})}_{i, - {n \over 2}}(g) $, in agreement with
(\ref{pint9}).

We also have an explicit formula for $d\mu \,\rho$. Notice that
the integral of $d\mu \,\rho$ is $N$, the number of states or the dimension
of the LLL. Since they are the kernel of the antiholomorphic derivative as
in (\ref{matt5b}), they are given by the integral of the Dolbeault
index density. The appropriate formula for two dimensions is
\beq
\I_{\rm Dolb} =  i\,\left ({F \over 2 \pi} + {R \over 4 \pi}\right)
\label{matt13a}
\eeq
The $U(1)$ background gauge field we have chosen is
\beq
F = - i n \, \Omega_K
\label{matt15}
\eeq
where the normalization is specified as
$\int \Omega_K/(2\pi) = 1$.
The curvature of $S^2$ is given by
$R = -i \, 2\, \Omega_K$ and it is easily verified that $\I_{\rm Dolb}$ integrates
to $N = n+1$. We may thus expect that, even at the level of the density, before integration,
 $d\mu \, \rho$ can be identified as
the two-form $\I_{\rm Dolb}$,
\beq
d\mu\, \rho = \I_{\rm Dolb} = 
i\,\left ({F \over 2 \pi} + {R \over 4 \pi}\right)
\label{matt15a}
\eeq
 This is confirmed by several independent 
arguments.
The simplest way is to note that $\sum_i u_i^* u_i$ is the number density
of the fermions in the fully filled LLL.
This is essentially the charge density
and so it 
 is related to an effective action for the background
fields as
\beq
\delta S_{\rm eff} = \int d\mu(g) \,\rho \, (i \delta A_0) 
\label{matt24}
\eeq
where $A_0$ is the time-component of the background gauge potential.
(We take this to be antihermitian to agree with the convention for the other
fields.)
It is well known that the effective action is of the form
$- \int A(F +R) /(4\pi)$, as 
calculated by a number of authors, even allowing for variations from
the fixed background values of $F = -i n \Omega_K $, $R = - i 2\,\Omega_K$
\cite{{QHE1a},{QHE2}}.
The result (\ref{matt15a}) is then straightforward.
It is also easy to understand this intuitively, at least for the $U(1)$ background.
A change of the field by $\delta A$ is equivalent to the change of
symplectic structure as
$\Omega \rightarrow \Omega + d (i \delta A)$. The volume element of phase space
is then $i F = \Omega + d (i \delta A)$ and hence it is the appropriate density for
the number of states in the semiclassical approximation.
To recapitulate, the advantage of writing $d\mu\, \rho$  in terms of the index density
 is that it applies even with perturbations
to the gauge field or the geometry, so long as we remain within the same topological
class.

We can now combine this with the path-integral from (\ref{pint8b}).
The exponent in the path-integral is the symbol for the Poincar\'e-Cartan form
$\A$ of (\ref{pint8a}).
Consider the effective action in terms of the gauge field $A$ and
the spin connection $\omega$,
\beq
S_{\rm eff} (A, \omega ) = - {1\over 4\pi} \int ( A dA + A \,R )
\label{matt26}
\eeq
with $R = d  \omega$.
The path-integral is then given by
\beq
Z = {\cal N} \int [D \Pi \, D \phi] \, \exp \left( i S_{\rm eff} (A+ \A, \omega ) \right)
\label{matt27}
\eeq
Strictly speaking, we should use $S_{\rm eff} (A+ \A, \omega ) - S_{\rm eff} (A, \omega )$,
but the extra factor $e^{i S_{\rm eff}(A, \omega )}$ is a constant as far
as the integration over the matter fields is concerned and can be absorbed in the normalization factor
for now. In fact, there is good reason to keep $S_{\rm eff}( A, \omega )$
in (\ref{matt27}), it will be relevant for the dynamics of gravity itself.

Turning to higher dimensions, consider as an example, $\mathbb{CP}^2\times \mathbb{R}$.
The fuzzy version of $\mathbb{CP}^2$ can be modeled by a Hilbert space
$\H_N$  which can be identified as the LLL of a  Landau-Hall problem
on $\mathbb{CP}^2 \sim (SU(3)/U(2)$. 
One can choose a constant $U(2)$ background for the gauge field,
proportional to the curvatures of $\mathbb{CP}^2$. The wave functions are 
coherent states
or the holomorphic sections of a suitable line or vector bundle of the form
$\sqrt{N} \,\D^{(r)}_{k, w}(g) = \sqrt{N} \bra{r, k} g \ket{r, w}$ which is the matrix representative of an $SU(3)$ element
$g$
in the representation labeled as $r$. The state $\ket{r, w} $ has to be chosen to ensure
that the commutators of right translation operators on $\mathbb{CP}^2$
reproduce the chosen background field strengths. The Dolbeault index density takes the
form
\beq
\I_{\rm Dolb} = {1\over 2} \left( {i F \over 2 \pi} + { i R^0\over 2 \pi}\right)^2
- {1\over 12}\left[  \left( {i R^0 \over 2 \pi} \right)^2 + {1\over 2} \Tr \left( {i {\bar R}\over 2 \pi}
{i {\bar R}\over 2 \pi}\right) \right]
\label{matt28}
\eeq
where $R^0 = d \omega^0$ and ${\bar R} = -i t_a R^a$ are the $U(1)$ and $SU(2)$ curvatures, 
$t_a = {\half} \sigma_a$. The path-integral for matter fields takes the same form as
(\ref{matt27}), namely as the integral of $\exp( i S_{\rm eff}(A+\A, \omega) )$,
with $S_{\rm eff}$ given by \cite{KN-Dolb}
\beqar
S_{\rm eff} (A, \omega) &=&{i^3 \over {(2\pi)^2}} \int \Biggl\{ { 1 \over 3!} \bigl(A+ \omega^0\bigr) \Bigl[d\bigl(A+\omega^0\bigr)\Bigr]^2\nonumber\\
&&\hskip .7in -{ 1 \over 12} \bigl(A+ \omega^0\bigr) \Bigl[  (d\omega^0)^2 + {1\over 2} \Tr ( {\bar R} \wedge {\bar R})\Bigr] \Biggr\}
\label{matt29}
\eeqar
The Poincar\'e-Cartan form should also be defined with the star products appropriate to
fuzzy $\mathbb{CP}^2$.

What has emerged from the arguments presented in this section is a simple prescription on how to couple the
matter fields to the spatial fields or the gravitational  background at the level of the effective action,
namely, as in (\ref{matt27}). The $U(1)$ gauge field in $S_{\rm eff} (A, \omega)$
is shifted by the Poincar\'e-Cartan form $\A$ as $A \rightarrow A+ \A$
and the functional integration is done over $\Pi$, $\phi$.
This is the key result of the
analysis.
For the rest of this paper, we will explore possible generalizations.

%%%%%%%%%%%%%%%%%%%%%%%%%%%%%%%%%%%%%%%%%%%%%%%
%%%%%%%%%%%%%%%%%%%%%%%%%%%%%%%%%%%%%%%%%%%%%%%
\section{Generalizing the background geometry}
%%%%%%%%%%%%%%%%%%%%%%%%%%%%%%%%%%%%%%%%%%%%%%%
%%%%%%%%%%%%%%%%%%%%%%%%%%%%%%%%%%%%%%%%%%%%%%%

First, we want to consider (\ref{matt27}) in the context of three-dimensional gravity, starting
at the level of the effective action. It has been known for a long time that the 
gravitational action in (2+1) dimensions can be written as the integral of 
the difference of two Chern-Simons terms.
Since this requires the consideration of more general backgrounds, we first consider
a simplification of the matter field dynamics.
From what has been discussed before, 
the integral of the Poincar\'e-Cartan form for a scalar field has the structure
\beq
- \int d \mu \rho\,  \left[ dt \,\Pi* {\dot \phi}  - dt H \right]
= {1\over 4 \pi} \int (2 F + R)\,\left[ dt \,\Pi* {\dot \phi}  - dt H \right]
\label{matt33}
\eeq
where $H$ is as given in (\ref{pint8}).
So far we have used dimensionless coordinates, normalizing the volume of
$S^2$ to 1. We restore the normal assignment of dimensions by the scaling
\beq
dx \rightarrow { d x \over a l}
\label{matt33a}
\eeq
where $l$ has the dimensions of length and $a$ is a constant, to be fixed shortly.
On a background such as $S^2\sim \mathbb{CP}^1$ with $F = -i n\, \Omega_K$, and the
K\"ahler form is normalized such that $\int \Omega_K /(2 \pi) = 1$, we get
\beq
d \mu \rho =  n \, i {dz d\bz \over (1 + \bz z )^2} + R\mhyphen{\rm term}
\longrightarrow {n \over a^2 l^2} i {d^2x \over (1 + x^2 /(a^2 l^2))^2}
\label{matt34}
\eeq
We choose $a^2 = n$ now and also define $r = a l$ as the radius of the sphere.
Then
\beq
- \int d \mu \rho\,  \left[ dt \,\Pi* {\dot \phi}  - dt H \right]
\rightarrow \int dt {d^2x \over (1+ x^2/r^2)^2}~
\left[ {\Pi*{\dot \phi} \over l^2} - {H \over l^2} \right]
\label{matt35}
\eeq
We now introduce a further scaling of the fields by writing
$\Pi = l \, {\tilde \Pi}$, 
$\phi = l\, {\tilde\phi}$.
The first term on the right hand side
of (\ref{matt35}) becomes $\int dV {\tilde \Pi}\, {\dot{\tilde \phi}}$, where
$dV$ denotes the volume element. The integral of the Hamiltonian becomes
\beq
\int dt \,H  \rightarrow  \int dV \, \left[ {1\over 2} \left( {\tilde \Pi}* {\tilde \Pi} 
+ \alpha_1 a^2 l^2 (\nabla_y{\tilde \phi} ) * (\nabla_y {\tilde \phi}) 
+ m_0^2\, {\tilde \phi}*{\tilde \phi} \right) + \lambda_0 l^2 {\tilde \phi}*{\tilde \phi} *{\tilde \phi}*{\tilde \phi} 
\right]
\label{matt36}
\eeq
We can now choose $\alpha_1 a^2 l^2 = 1$ to set the spatial gradient term to the usual
form. (This is equivalent to choosing a speed of light.)
Further we have to identify the
$m_0$ and $\lambda_0 l^2$ as the new bare mass and bare coupling constant. The
Hamiltonian then takes the
standard form, but with star products.

Let us now consider the higher terms in the star product involving gradients of the fields.
 The first corrections are of the
form
\beq
 {R_+ f \, R_- h \over n} \sim  l^2 {R_+ {\tilde f}  \, R_- {\tilde h} \over n}
\label{matt37}
\eeq
where $f$, $h$ could be $\Pi$ or $\phi$.
The derivatives, as written, are dimensionless. After the scaling of coordinates
as in (\ref{matt33a}), this takes the form
\beq
l^2 {R_+ {\tilde f} \, R_- {\tilde h} \over n} \sim (\nabla_y {\tilde f}) (\nabla_y {\tilde h}) { a^2 l^4 \over n}
\sim { (\nabla_y {\tilde f}) (\nabla_y {\tilde h})\over M^4}
\label{matt38}
\eeq
where $M = l^{-1}$. Notice that, so far, the scale of $M$ is not fixed by anything.
So there is some freedom in choosing this.
Since the number of states is $n$ (at large $n$) and the volume of the spatial
universe is $ a^2 l^2 = n l^2$, we see that $l$ or $M^{-1}$ 
determines the size of one elemental state for the spatial manifold.\footnote{If we interpret this within a gravity theory, the Planck scale is a natural choice for this.
But $M$ could be somewhat smaller or larger, although in any realistic sense,
$M$ cannot be too low, since it determines the limit of resolution
for points of the spatial manifold itself.}
The corrections from the star products are therefore negligible in a regime of energies low
compared to $M$ when the magnitudes of $\nabla \Pi /M^2$ and
$\nabla \phi /M^2$ are small.
In this limit, we can replace the Poincar\'e-Cartan form 
by
\beq
\A =  -i\, \left[ \Pi\, {\dot \phi}  - dt H \right]\,dt
\label{matt39}
\eeq
where star products are neglected in the expression for $H$ as well.
This is a significant simplification which is helpful for generalization, for the following reason.
The star product is specific to a particular background. Although it can be generalized to some
extent, it is tied to having holomorphicity for the wave functions used to construct
the operators from the contravariant symbols.\footnote{It is possible to define a star product for any Poisson manifold \cite{Kont}. But it is not clear how to use it in the present context.}
This is an obstruction to the
framing of the present problem within the context of general gravitational backgrounds.
However, for the simplified version in (\ref{matt39}), it
 can be done if we are interested in low energy dynamics for the matter fields
 where the star products are
not important.

On a general gravitational background (which does not necessarily have a complex structure)
we cannot use the Dolbeault index density, rather we shall consider the Dirac
index density.
Our aim is to show that the Chern-Simons form associated with the
Dirac index density will reduce to the effective actions (\ref{matt26}) and
(\ref{matt29}) when a particular  choice of background is made.
Therefore, the Chern-Simons forms serve as the effective action to be used
in (\ref{matt27}) for a general gauge and gravitational background.
 
Towards showing this result, for the $2+1$ dimensional case, we start with the Dirac index density
in four dimensions which is given by
\beqar
\I_{\rm Dirac} &=& - {{\rm dim}V \over 24} \,p_1  - {1\over 2} \Tr \left(
{F \over 2 \pi} {F \over 2 \pi} \right)\nonumber\\
&=&\left[  {{\rm dim}V \over 48} 
\Tr \left( {R \over 2 \pi} {R \over 2 \pi} \right) - {1\over 2} \Tr \left(
{F \over 2 \pi} {F \over 2 \pi} \right)\right]
\label{ind1}
\eeqar
In the second line of this equation, the
 curvatures are written in terms of the vector representation of 
$SO(4)$ so that $\Tr (R R) = R^{ab} R^{ba}$, $a, b = 1,2,3,4$;
$p_1$ in the first line
is the Pontrjagin class given by $p_1 = R^{ab} R^{ab} /(8\pi^2)$.
Also ${\rm dim}V$ is the dimension
of the vector bundle or the dimension of $F$'s viewed as matrices.
For an Abelian background field, which is our case, ${\rm dim}V = 1$. 
The Chern-Simons term corresponding to (\ref{ind1}) is given by\footnote{Our normalization is
$ d (C.S.) = 2\pi \times$(Index density).}
\beq
S_{\rm eff} = \int  \left[  {1\over 96 \pi} 
\Tr \left(\omega\, d\omega + {2\over 3} \omega^3 \right) - {1\over 4 \pi} 
A \,dA\right]
\label{ind2}
\eeq

We want to argue that (\ref{ind2})  is the effective action (or at least part of it)
for our problem on a general gauge and gravitational background.
Towards this, we 
will now show that this does lead to (\ref{matt27}) if we take the spacetime manifold to
be $S^2 \times \mathbb{R}$.
In this case, the spin connection has only the nonzero
component $\omega^{ij} = i \epsilon^{ij} \,\omega$, defined by the zero torsion condition
$de^i + \omega^{ij} e^j = 0$, where $e^i = (e^1, e^2)$ are the frame fields for
$S^2$. ($e^3 = dt$ will be the third frame field, for $\mathbb{R}$.)
The action (\ref{ind2}) now reduces to
\beq
S_{\rm eff} = \int  \left[  {1\over 48 \pi} 
\,\omega\, d\omega - {1\over 4 \pi} 
A \,dA\right]
\label{ind3}
\eeq
To compare this with the effective action for $S^2 \times \mathbb{R}$ as in
(\ref{matt26}), (\ref{matt27}), two changes are needed. Here we are discussing
spinors, while (\ref{matt26}), (\ref{matt27}) apply to scalars where we could use the
Dolbeault index density. Spinors transform nontrivially as $\psi \rightarrow e^{i \sigma_3 \vf/2} \psi$ under local spatial
rotations while scalars do not respond to rotations. So the factor $e^{i \sigma_3 \vf/2}$
must be canceled out to get a proper comparison with the Dolbeault index
density. This can be done by the shift $A\rightarrow A+ \half \omega$  in (\ref{ind3}).
(In other words, we can view the case of scalars as this particular choice of 
background fields for the spinors.)
And secondly, we must make the replacement $A \rightarrow a +\A$
to include the matter coupling. With these changes, the action becomes
 \beq
 S_{\rm eff} ( A +  \A + {\half}  \omega, \omega)
 = - {1\over 4 \pi} \int \left[ A dA + A R + {1\over 6} \omega d\omega\right] - \int \A \left( {F \over 2 \pi} + {R \over 4 \pi}
 \right) 
 \label{ind4}
 \eeq
 The first set of terms agrees with the action obtained in \cite{KN-Dolb} and the second set of terms
 agrees with the present discussion in (\ref{matt24})-(\ref{matt27}). Thus we have shown that
the general action (\ref{ind2}) reduces to (\ref{ind4}) so that
our earlier results for $S^2 \times \mathbb{R}$ can be viewed as a special choice of
background fields. Returning to the general case, we see that the 
action for describing the path-integral for
matter fields takes the form of (\ref{matt27}) with
\beqar
S_{\rm eff} &=& \int  \left[  {1\over 96 \pi} 
\Tr \left(\omega\, d\omega + {2\over 3} \omega^3 \right) - {1\over 4 \pi} 
(A+ \A) \,d(A +\A) \right]\nonumber\\
&=& \int  \left[  {1\over 96 \pi} 
\Tr \left(\omega\, d\omega + {2\over 3} \omega^3 \right) - {1\over 4 \pi} 
A\,d A  \right] - {1\over 2 \pi} \int \A \, dA
\label{ind5}
\eeqar

Turning to the $4+1$-dimensional case, the Dirac index density in six dimensions
is given by
\beq
\I_{\rm Dirac} = {{\rm dim} V}\, {i F \over 2 \pi}
\left[ {1\over 3!} \left( {i F \over 2 \pi} \right)^2 - {1\over 24} p_1\right]
+ {1\over 2} \left({i F \over 2 \pi} \right) \,\Tr \left( { i {\bar F}\over 2 \pi}
\,{i {\bar F}\over 2 \pi}\right) 
\label{ind6}
\eeq
Here $F$ is the $U(1)$ field strength, ${\bar F}$ is the $SU(2)$ background field
and ${\rm dim}V$ is the dimension of the representation used for ${\bar F}$.
As before, $p_1$ is the Pontrjagin class.
The Chern-Simons form or effective action corresponding to (\ref{ind6})
is given by
\beq
S_{\rm eff} (A, \omega ) = {i^3 {\rm dim}V}
\int \left[ {1\over 3! (2 \pi)^2} A F F 
+  {1\over 24} A \, p_1 \right]
+ {i^3\over 8\pi^2} A\, \Tr ({\bar F} {\bar F} )
\label{ind7}
\eeq

Again, our first task will be to show that this will lead  to the effective action 
(\ref{matt29}) when we make a special choice of background fields.
We consider the case of zero nonabelian background, i.e., ${\bar F} = 0$, 
${\rm dim}V = 1$. 
In reducing (\ref{ind7}) to spatial states defined by the Landau-Hall problem for
scalars
on a background geometry of the form $\mathbb{CP}^2 \times \mathbb{R}$, 
there are two requirements on the $U(1)$ field. 
First of all, since $\mathbb{CP}^2$ does not
admit a spin structure, the use of the Dirac index density is problematic.
One can use a spin$^c$ structure, which means that
we should make a shift of the $U(1)$ as $A \rightarrow A + {\half} \omega^0$.
(This is equivalent to choosing a $U(1)$ charge of the form $n+{\half}$
where $n$ is an integer.) This shift reduces the problem to
spinors on $\mathbb{CP}^2$ (with a spin$^c$ structure).
(For a discussion of the Dirac index for $\mathbb{CP}^2$, relevant for our
analysis, see \cite{Dolan}.)

To get to scalars, so that we can compare with the Dolbeault index, we need a further shift
by $\half \omega^0$ to compensate for the transformation of spinors under rotations.
Thus, in total, we should use $A \rightarrow A + \omega^0$. Further, in terms of the $U(1)$ and
$SU(2)$ curvatures,
$p_1$ reduces to
\beq
{p_1 \over 24} = {1\over (2 \pi )^2} \left[ - {1\over 12} d\omega^0\, d\omega^0 - {1\over 24} \Tr
({\bar R} {\bar R}) \right]
\label{ind8}
\eeq
With $A \rightarrow A + \omega^0$ and this formula for $p_1$, the effective action
(\ref{ind7}) 
becomes
\beq
S_{\rm eff} = {i^3 \over (2\pi)^2}\int \Biggl[
{1 \over 3!} (A + \omega^0) (d A + d \omega^0)^2 
- {1\over 12}  (A + \omega^0) \Bigl[ d \omega^0\, d\omega^0 + {1\over 2}
\Tr ( {\bar R} {\bar R} ) \Bigr] \Biggr]
\label{ind9}
\eeq
This agrees with the effective action obtained in \cite{KN-Dolb} using the Dolbeault
 index theorem and quoted in (\ref{matt29}).
 The charge density which is the variation of
 $S_{\rm eff}$ with respect to $A_0$ gives the correct $d\mu \rho$ for this case,
 and so the prescription (\ref{matt27}) for fuzzy $\mathbb{CP}^2$ is recovered for
 the particular choice of background.

Having shown that (\ref{ind7}) does indeed reproduce the results for
$\mathbb{CP}^2 \times \mathbb{R}$, we can take it as the form of the
action  for general, not necessarily complex K\"ahler,
backgrounds. The shift by the Poincar\'e-Cartan form produces the
matter part of the action
\beq
S_{\rm matter} = {1\over 32\pi^2}\int (i \A) 
\Bigl[ {\rm dim}V \Bigl(F_{\mu\nu} F_{\alpha \beta} 
+ {1\over 24} R^{ab}_{\mu\nu} R^{ab}_{\alpha\beta}  \Bigr)
+ \Tr (t_a t_b) {\bar F}^a_{\mu\nu} {\bar F}^b_{\alpha\beta})\Bigr] dx^\mu\cdots dx^\beta
\label{ind10}
\eeq
where we used the real field components defined by
\beq
F = (-i )\, {\half} F_{\mu\nu} dx^\mu \,dx^\nu , \hskip .2in
{\bar F} =  (-i t_a)\,{\half} F^a_{\mu\nu} dx^\mu \,dx^\nu
\label{ind11}
\eeq
The key emerging feature is that the Lagrangian for matter fields is multiplied by a specific polynomial
involving powers of the curvature.
The term involving the $U(1)$ field, namely,
$F_{\mu\nu} F_{\alpha \beta}$ is the dominant one at large $n$, but there are curvature-dependent subdominant terms. It is curious to note that such couplings of matter fields to gravity
have been extensively investigated recently, partly motivated by their potential
to explain observations related to
dark matter. For recent reviews on the subject of curvature-matter couplings,
 see \cite{dark}.

Finally we can ask how to formulate the coupling of matter fields directly at the level of 
the fermion fields defining the spatial geometry.
The relevant action is the Dirac action with nontrivial gauge and gravitational fields.
The gauge group should have a $U(1)$ component. For example, in 4+1 dimensions, we
consider the action
\beq
S = \int {\bar \psi} (i \gamma\cdot D ) \psi
\label{ind12}
\eeq
If a specific representation of the Dirac matrices is needed, we will use
\beq
\gamma^0 = \left( \begin{matrix}
1&0\\ 0&-1\\ \end{matrix} \right), \hskip .2in
\gamma^i = \left( \begin{matrix}
0&\sigma^i\\ -\sigma^i&0\\ \end{matrix} \right), \hskip .2in
\gamma^4 = \left( \begin{matrix}
0&i\\ i&0\\ \end{matrix} \right)
\label{ind13}
\eeq
In (\ref{ind12}), we are using a 4-component spinor $\psi$ which would correspond to one chiral component
of an 8-spinor in six dimensions. This means that there will be a parity anomaly for
this theory. The Hamiltonian corresponding to (\ref{ind12}) is
\beq
H = \int \psi^\dagger \left[ - i \gamma^0 \gamma^\mu D_\mu \right] \psi ,
\hskip .2in \mu = 1,2,3,4.
\label{ind14}
\eeq
The nonzero positive and negative eigenvalues of the 4-d Dirac operator
$- i \gamma^0 \gamma^\mu D_\mu $ are paired, with the corresponding eigenfunctions
$\psi_n$ and $ \gamma^0\psi_n$. The zero modes are not paired 
and their number is what is given by the Dirac index in four dimensions.
In defining the vacuum state and calculating the charge, the key question is whether the zero
modes are to be considered as part of the Dirac sea, hence filled, or as part of the
unoccupied positive energy states. 
The charge conjugation transformation for the spinor $\psi$ is defined by
\beq
\psi = C \phi^*, \hskip .2in C^{-1} \gamma^a  C = \gamma^{a*},
\hskip .2in C = \gamma^2 \gamma^4
\label{ind15}
\eeq
where $\phi$ is the charge conjugate of $\psi$. The $C$-odd definition of the
charge can be evaluated on the vacuum as
\beq
Q \, \ket{0_\mp} = {1\over 2} \int \left[ \psi^\dagger \psi  - \phi^\dagger \phi \right]
\, \ket{0_\mp} =  \mp {\half} \, N \ket{0_\mp}
\label{ind16}
\eeq
where the upper sign corresponds to the zero modes being unoccupied, the lower to
the case when they are occupied and $N$ is the number of zero modes given by the Dirac
index.
 This result shows that the effective action for
(\ref{ind12}) should have a Chern-Simons term with a level number of $\mp \half$.
This leads to an inconsistency. (This is the well known parity anomaly, spelt out for 4+1 dimensions here.)
A consistent theory requires using the Dirac action (\ref{ind12}) with
a Chern-Simons term with level number $\pm \half$ added. Taking the first sign in (\ref{ind16}),
the resulting vacuum will have zero charge and will lead to an effective action equal to
the Chern-Simons term in (\ref{matt29}) for the fully occupied $\nu =1$ state, i.e., 
for the state where all the zero modes are occupied.
Thus our results for the coupling of matter fields
to the fermions characterizing the spatial manifold are summarized by the action
\beqar
S &=& \int {\bar \psi} (i \gamma\cdot D - m  ) \psi
+ {1\over 2} S_{CS}(A + \A, \omega )\nonumber\\
\gamma\cdot D &=& \gamma^a e^{-1\mu}_a \left(\del_\mu + A_\mu + \A _\mu + {1\over 8 i} \omega^{bc}_\mu [\gamma_b, \gamma_c] \right)
\label{ind17}
\eeqar
The mass $m$ is a small positive number whose role is to shift the energies upward.
The zero modes of the Hamiltonian (\ref{ind14}) will thus have small
positive energies, leaving them unoccupied and making the choice of the vacuum state
as $\ket{0_-}$ in (\ref{ind16}).
 (This is the only reason for $m$; it can be taken to
be infinitesimally small.)
And $S_{CS}$ in (\ref{ind17}) is
the Chern-Simons action of (\ref{ind7}) with the shift
$A \rightarrow A +\A$,
\beq
S_{CS}(A+ \A ,\omega ) = \Biggl[{i^3 {\rm dim}V}
\int \left[ {1\over 3! (2 \pi)^2} A F F 
+  {1\over 24} A \, p_1 \right]
+ {i^3\over 8\pi^2} A\, \Tr ({\bar F} {\bar F} )\Biggr]_{A\rightarrow A +\A }
\label{ind18}
\eeq
The effective action, for the state with the zero modes
fully occupied,
 obtained from (\ref{ind17}) will be $S_{CS}$ as in (\ref{ind18}).

%%%%%%%%%%%%%%%%%%%%%%%%%%%%%%%%%%%%%%%%%%%%%%%
%%%%%%%%%%%%%%%%%%%%%%%%%%%%%%%%%%%%%%%%%%%%%%%
\section{Discussion}
%%%%%%%%%%%%%%%%%%%%%%%%%%%%%%%%%%%%%%%%%%%%%%%
%%%%%%%%%%%%%%%%%%%%%%%%%%%%%%%%%%%%%%%%%%%%%%%
Our analysis started with a finite, say $N$, dimensional Hilbert space of states which could
be identified as  the lowest Landau level  of a Landau-Hall problem or as the Hilbert space
modeling a fuzzy space. Observables on such a space are $N \times N$ matrices.
We considered the path-integral for the dynamics of such observables, specifically something which
approaches a continuum field theory as $N \rightarrow \infty$.
The action which defines such a path-integral is given by
a Chern-Simons form which includes a $U(1)$ gauge field $A$ which is shifted as
$A \rightarrow A + \A$ by the (star product version of the) Poincar\'e-Cartan form
$\A$ for the matter fields. We then extended this to more general backgrounds
arguing that the Dirac index density can be used to construct the relevant
Chern-Simons form. As far as matter fields are concerned, the end result is an action of the form
$\int \rho\, \L$ where $\L$ is the Lagrangian and $\rho$ is a density which is a polynomial
in the gauge fields and the curvature as determined by the index density.
It was argued in \cite{KN-Dolb} that the effective action for background gauge fields and gravity
for the Landau-Hall system is given by
a Chern-Simons form associated with the Dolbeault index density. The present work
incorporates matter couplings in such a framework and further extends it to more general
geometries.

As mentioned after (\ref{matt27}), $S_{\rm eff}(A+ \A,\omega )$ has a
term $S_{\rm eff}(A,\omega )$ which is not related to matter couplings. We retained it in
(\ref{matt27}) expecting that such terms can be absorbed into the gravitational part 
of the action. Regarding such purely gravitational terms, we note that one
can define a class of gravity theories on odd-dimensional spacetimes
with an action which is the difference of two Chern-Simons forms. We have argued 
elsewhere for the natural
emergence of such a structure with an interpretation in the framework
of thermofield dynamics \cite{VPN1}. The inclusion of matter couplings as discussed here within such a
structure would be an interesting next step, which we propose to pursue in a later publication.

 \bigskip

I thank Dimitra Karabali for a careful reading of the manuscript and
for useful comments.
This research was supported in part by the U.S.\ National Science
Foundation grant PHY-1820721
and by PSC-CUNY awards.
 
%%%%%%%%%%%%%%%%%%%%%%%%%%%%%%%%%%%%%%%%%%%%%%%
%%%%%%%%%%%%%%%%%%%%%%%%%%%%%%%%%%%%%%%%%%%%%%%

%%%%%%%%%%%%%%%%%%%%%%%%%%%%%%%%%%%%%%%%%%%%%%%
%%%%%%%%%%%%%%%%%%%%%%%%%%%%%%%%%%%%%%%%%%%%%%%
%%%%%%%%%%%%%%%%%%%%%%%%%%%%%%%%%%%%%%%%%%%%%%%
%%%%%%%%%%%%%%%%%%%%%%%%%%%%%%%%%%%%%%%%%%%%%%%
\end{document}